# Improving visibility in photoacoustic imaging using dynamic speckle illumination

Jérôme Gateau, Thomas Chaigne, Ori Katz, Sylvain Gigan and Emmanuel Bossy*

*Institut Langevin, ESPCI ParisTech, CNRS UMR 7587, INSERM U979, 1 rue Jussieu, 75005 Paris, France*
*Corresponding author: emmanuel.bossy@espci.fr*

In high-frequency photoacoustic imaging with uniform illumination, homogeneous photo-absorbing structures may be invisible because of their large size or limited-view issues. Here we show that, by exploiting dynamic speckle illumination, it is possible to reveal features which are normally invisible with a photoacoustic system comprised of a 20MHz linear ultrasound array. We demonstrate imaging of a Ø5 mm absorbing cylinder and a 30 µm black thread arranged in a complex shape. The hidden structures are directly retrieved from photoacoustic images recorded for different random speckle illuminations of the phantoms by assessing the variation in the value of each pixel over the illumination patterns.

In biomedical imaging, randomly distributed sub-resolution sources or scatterers usually result in speckle artifacts. In ultrasonography, acoustic speckle is of primary importance for characterization of soft tissue [1] as it enables the visualization of large or complex-shaped structures comprised of unresolved scatterers, even when they are densely packed. On the other hand, photoacoustic imaging is known to be mostly speckle-free for uniform illumination and large density of optical absorbers contained in structures with smooth boundaries [2]; blood vessels filled with hemoglobin for instance. The lack of acoustic speckle artifacts in photoacoustics has been attributed to the strong initial phase and amplitude correlation among the ultrasound waves generated by the individual absorbing molecules or particles after quasi-instantaneous optical excitation. Two related effects of this correlation are the directivity of the ultrasonic emission for elongated structures, and the prominent boundary build-up for large structures. This results in visibility issues when imaging homogeneous structures using a photoacoustic system with uniform illumination, because of the limited bandwidth and/or the limited view of practical implementations [3].

To address this visibility problem, we propose here to exploit optical speckle patterns naturally present in coherent illumination as a source of structured illumination, effectively breaking the homogeneity of absorbing structures and rendering them visible in photoacoustic imaging. Dynamic speckle illumination has proved very useful in optical microscopy for optical sectioning [4], and was also recently exploited as a structured illumination source to surpass the optical diffraction limit [5]. As photoacoustic techniques aim at imaging beyond the depth achieved by optical microscopy [6], only random a-priori unknown structured illumination provided by the passage of coherent light through a scattering medium can be considered.

In the present study, unlike in conventional photoacoustic imaging where a locally uniform illumination is assumed, we investigate the operation of a high frequency and limited-view photoacoustic system with dynamic optical speckle illumination of homogeneously absorbing structures. The final images are obtained by computing the local variations of the signal under the different speckle illumination.

The experimental set-up is illustrated in Fig. 1(a). Optical excitation was performed with a Q-Switched Nd:YAG oscillator laser (Brillant, Quantel) delivering 4 ns duration pulses at $\lambda_{laser}$=532 nm with a 10 Hz repetition rate, and a coherence length on the order of 1 mm. To generate the varying speckle illumination, the laser beam was passed through a ground glass diffuser (220 Grit, Thorlabs), positionned on a rotation mount. The rotating diffuser acted here as a random inhomogeneous dynamic medium producing temporally varying speckle patterns with no appreciable ballistic transmitted component; just like a thick scattering biological tissue would. The main difference is the scattering-induced path-length differences. Path-length differences larger than the laser coherence length may reduce speckle contrast, and were avoided here by employing a thin scattering layer.

The diffuser was positioned at a distance L=15 mm to 150 mm from the imaged plane. Varying this distance allowed to control the optical speckle grain size on the target plane [7]. In our experiments L was varied to produce transverse speckle grain sizes ranging from $D_{speckle}$ = 2.6 µm ± 0.6 µm to 18.3 µm ± 1.7 µm. $D_{speckle}$ was determined prior to the photoacoustic experiments by imaging the optical field on a camera (Manta G-046, Allied Vision Technologies) [Fig. 1(b)], and calculating the $1/e^2$ radius of the normalized spatial autocorrelation function of the optical speckle image.

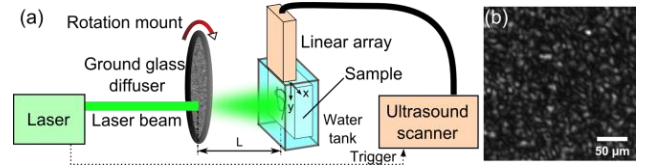

Fig.1: (a) Schematic of the experimental setup. The laser input beam (diameter ~ 6 mm) impinged on the diffuser surface, which was placed at a distance L of the imaging plane. A coordinate system is introduced for the imaging plane. Its origin is the middle point of the array. (b) A typical optical speckle pattern recorded with the camera in the xy plane (transverse grain size of 6.3 µm ± 0.8 µm).

Two samples were prepared to mimic the most common artifacts in high-frequency photoacoustic imaging: boundary build-up [2] [Fig. 2] and limited view [8] [Fig.3], and to verify the benefits of our technique. Both samples were imaging phantoms comprised of homogeneous photo-absorbing structures embedded in 1.5%w/v agarose gel (A9539, Sigma). The gel supporting the sample was weakly optically scattering, and did not affect significantly the incident speckle. The gel also mimicked the speed of sound in biological tissue. The first phantom (Phantom 1) contained a cylinder of 5 mm in diameter and 3 mm thick prepared by mixing diluted black ink (Scribtol, Pelikan) with agarose gel to obtain an optical density of 1.5 [Fig 2(a)]. The second phantom (Phantom 2) comprised a 30 µm diameter black-colored nylon suture thread (NYL02DS, 10/0, Vetsuture) arranged in a two-loop knot [Fig 3(a)].

Detection of the photoacoustic waves was performed with a 128-element linear array (Vermon, France) driven by a programmable ultrasound scanner (Aixplorer, Supersonic Imagine, France). The elements of the array had a center frequency of 20 MHz. They were cylindrically focused at 8 mm and were distributed with a pitch of 80 µm. Water was used for acoustic coupling. Signals were acquired simultaneously on 128 channels at a sampling rate of 60 MS/s for every laser pulse. In each experiment, the measurement sequence consisted of the acquisition of signals corresponding to 50 consecutive laser pulses while continuously rotating the diffuser.

Following each acquisition, the signals were filtered with a 3rd order Butterworth band-pass filter between 500 kHz and 25 MHz, and a delay-and-sum beamforming algorithm was used to obtain a dataset of 50 photoacoustic images for the different speckle illuminations. To form the final images, for each pixel p in the image grid, two statistical quantities were estimated over the pixel values $p_i$ (i=1..N) corresponding to N different realizations of the speckle illumination: its average value and its variability. Specifically, the arithmetic mean µ(p) and the Gini mean difference (GMD) were computed. Because of the statistical properties of speckle illumination [7] and the linearity of the beamforming process, the limit of µ(p) when N tends toward infinity corresponds to the value obtained with a spatially uniform illumination. The statistical dispersion of the values $p_i$ over N realizations of the speckle illumination was estimated using the GMD [9]:

$$GMD_N = \frac{\sum_{i=1}^{N-1}\sum_{j=i+1}^{N}\left|p_i - p_j\right|}{N(N-1)/2} \quad (1)$$

The GMD has the advantage, over other estimators of the dispersion, such as the variance or standard-deviation, that the comparison is done between pairs and it is not defined in terms of a specific measure of central tendency. Moreover, the GMD gives equal weight to all the differences and for speckle illuminations can be computed even for N~2, when the variance of the mean illumination is large. Each photoacoustic reconstructed dataset was normalized by its global maximum before computing the mean image over N=50 realizations of speckle illumination, and the GMD images for N=2 and N=50. For better visualization of the images, the pixels with values below the most frequent value in the image (mode) were discarded, and the minimum value was then subtracted to the remaining pixels.

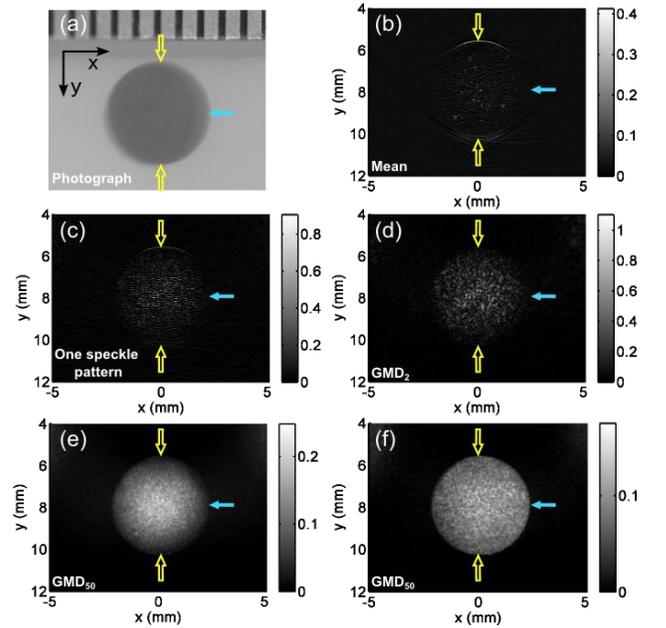

Fig. 2: Images obtained with Phantom 1. (a) Photograph of the Ø5 mm ink inclusion. A metric ruler was placed on the top for size reference. (b-e) Photoacoustic images from illumination with speckle patterns featuring a transverse grain size of $D_{speckle}$ = 2.6 µm ± 0.6 µm: (b) Mean image over 50 speckle patterns, mimicking a uniform illumination, (c) reconstruction from a single speckle-illumination pattern, (d) $GMD_2$ image, and (e) $GDM_{50}$ image. (f) same as (e) but for $D_{speckle}$ = 6.3 µm ± 0.8 µm. The superimposed empty and full arrows indicate surfaces respectively parallel and perpendicular to the length axis of the array. GMD images enable reconstruction of a large absorbing structure with a high-frequency photoacoustic system without artifacts.

Fig. 2(b)-2(f) illustrate photoacoustic images obtained with Phantom 1. Because of its shape, size and homogeneous absorption, the ink inclusion comprised in Phantom 1 [Fig. 2(a)] is expected to mostly emit ultrasound frequency in the sub-megahertz range when illuminated uniformly [10]. Such frequencies cannot be recorded efficiently with a high-frequency piezoelectric transducer. Therefore, the interior absorption could not be visualized on the mean image [Fig 2(b)], as it is the case in conventional photoacoustic imaging. Only the edges facing the detector (empty arrows) and a few sparse absorbers, most probably dust particles or large ink particles were visible. The side edges (full arrow) could not be reconstructed because of the limited aperture of the array along the x-axis. However, when illuminated with a speckle pattern, photoacoustic sources could be reconstructed inside the inclusion in addition to the edges [Fig 2(c)]. More sources inside can be visualized when calculating the $GMD_2$ image, obtained using only two different speckle illuminations [Fig. 2(d)]. This phenomenon can be explained by the heterogeneous spatial distribution of the light intensity, which breaks the amplitude correlation among the ultrasound waves generated by each point-like absorber throughout the structure, on a scale allowing the propagation of high frequency ultrasound components [Fig. 1(b)]. Assessing the variability of each pixel over 50 different speckle illuminations, the $GMD_{50}$ image [Fig. 2 (e)] clearly shows that the absorbing inclusion could be visualized and appeared with a uniform brightness, closely reproducing its absorption profile [Fig. 2(a)]. The edges are however not very sharp, but this phenomenon can be attributed to the lower average fluence in the periphery of the inclusion due to the small distance between the diffuser and the sample, that in turn reduced the signal-to-noise ratio for the corresponding pixels, and induced a dispersion similar to the background noise. Placing the diffuser further away increase the illuminated surface and enabled to restore sharp edges as well as the interior absorption [Fig. 2(f)]. Fig. 2(e) and 2(f) demonstrate that computing GMD images, even with the smallest speckle grain size tested here, enabled to reconstruct a large photo-absorbing structure with a high-frequency photoacoustic system.

Fig. 3(b)-3(f) illustrate photoacoustic images obtained with Phantom 2. The orientation of the black thread in this phantom was chosen to enhance the limited-view problem and to show how speckle illumination can restore visibility for all orientations. Three parts of the thread are visible on the mean image [Fig. 3(b)] (empty arrows): two portions mostly parallel to the x-axis and the top extremity of the thread. All other portions could not be distinguished on Fig. 3(b) and in particular the ones indicated with the full arrows. The images obtained with one speckle illumination [Fig. 3(c)] and $GMD_2$ [Fig. 3(d)] show that the portions of the thread indicated with the full arrows appear with a granular structure, as if they were formed of discrete absorbers and not a continuous one. However, with one speckle illumination, portions of the thread visible in the mean image (empty arrow) are predominant while all portions of the thread appear more homogeneously in the $GMD_2$ image. Computing the $GMD_{50}$ image [Fig 3(e)] restores the continuity of the photo-absorbing structure, and allows retrieving the complex shape of the structure [Fig 3(a)]. Similar results could be obtained with the smallest speckle grain size used in this study [Fig. 3(f)]; however as for Phantom 1, mainly the parts in the surrounding of the center of the image could be efficiently restored. Fig. 4(e) and 4(f) demonstrate that computing $GMD_{50}$ images enabled to reconstruct faithfully all the orientations of a photo-absorbing structure with a limited-view photoacoustic system. This effect results from the quasi omnidirectional radiation patterns of high-frequency photoacoustic waves when the speckle illumination was employed.

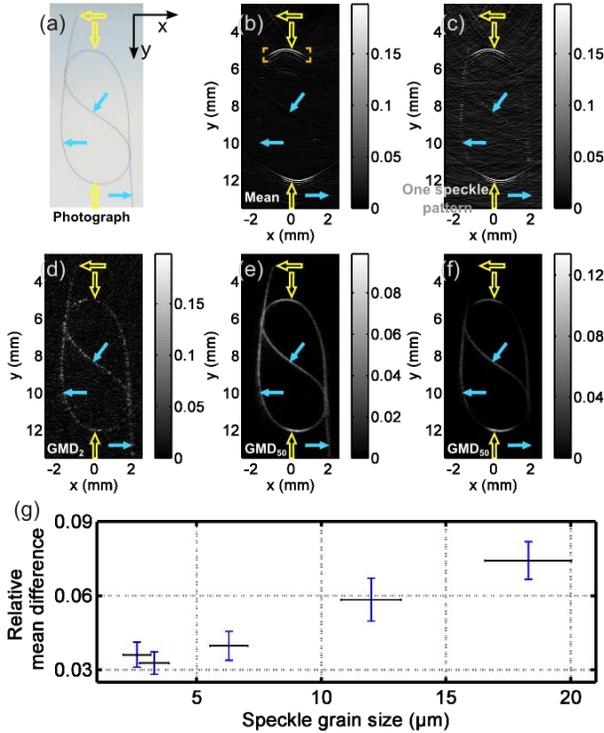

Fig. 3: Images obtained with Phantom 2. (a) Photograph of the Ø30 μm knotted thread. (b-e) Photoacoustic images from illumination with speckle pattern featuring a transverse grain size $D_{speckle}$ = 6.3 μm ± 0.8 μm: (b) Mean image over 50 speckle patterns, (c) Reconstruction from a single speckle-illumination pattern, (d) $GMD_2$ image from the illumination pattern in (c) and a second one. (e) $GDM_{50}$ image. (b) and (c) were threshold to the maximum value of (d). (f) same as (e) but for $D_{speckle}$ = 2.6 μm ± 0.6 μm. The superimposed empty and full arrows point out parts of the thread respectively mostly parallel and perpendicular to the length axis of the array. All the orientations of the thread can be retrieved in the GMD images. (g) Averaged RMD values -computed in the area delimited by the superimposed corners in (b)- as a function of the transverse speckle grain size. Vertical error bar: standard deviation, horizontal error bar: estimated precision on $D_{speckle}$ with the camera measurements.

To quantify the effect of the speckle grain size on the reconstructed images, the average variability in pixel value over the realizations of the speckle illumination was assessed, independently of the laser fluence on Phantom 2. This was done by estimating the relative mean difference (RMD) – or Gini ratio – for pixels with a large arithmetic mean µ (pixels with | µ | superior to 3 times the standard deviation of µ over all the pixels in the region of interest):

$$RMD = \frac{GMD_{50}}{2 \cdot |\mu|} \quad (2)$$

The RMD is the analogue of the optical speckle contrast [7] but computed per pixel, and conveys information over the volume that contributes to the signal within the pixel. We term this volume "acoustic cell". The averaged RMD value quantifies how different two photoacoustic images acquired for two different speckle illuminations are, and is related to the number of optical speckle grains illuminating optical absorbers in the acoustic cells [11]. A RMD of zero expresses perfect equivalence in the obtained signals between two uncorrelated speckle illuminations [9] and is expected to occur for sufficiently small speckle grain size compared to the acoustic cell [2]. A RMD of one expresses maximal inequality among values [9] and is only attainable for optical speckle grain equal or larger than the acoustic cell (i.e. speckle contrast equal to 1). Fig. 3(g) shows that the mean RMD value obtained for Phantom 2 increases with the optical speckle grain size but remain well below 1. Here, the transverse speckle grain sizes were at least three times smaller than the acoustic wavelength in water at the low-pass cut-off frequency (25MHz), therefore individual grains cannot be resolved. The two smallest speckle grains, i.e. 2.6 µm ± 0.6 µm and 3.3 µm ± 0.6 µm, have very similar mean RMD values. These speckle grains have transverse dimensions about 20 times smaller than the smallest measured acoustic wavelength and 10 times smaller than the structure diameter, but still allow image reconstruction.

To conclude, randomly structured illumination using speckle patterns produced by a dynamic scattering medium was demonstrated to compensate for the partial visibility issues in high-frequency photoacoustic imaging. The speckles were generated with a scattering layer at a distance from the sample to allow control of the experimental parameters and because of the relatively short coherence length of the laser employed. To implement the demonstrated technique inside scattering media, the coherence length of the laser should be longer than the scattering induced optical path-length differences. In addition, inside scattering tissues the speckle grain size is expected to decrease towards the quasi-constant value of $\lambda_{laser}/2$ within depths of the order of the transport mean-free path [12] (i.e. 1 mm for biological tissue in the near infrared [6]). The variability in each pixel decreases with the increasing number of optical speckle grains that contribute to this pixel. Therefore, higher acoustic resolution (smaller acoustic cell) provided by a higher upper cutoff frequency, may compensate for small optical speckle grain size at depth in tissue. Recently a photoacoustic system operating at a low-pass cut-off frequency of 125MHz and reaching depths of at least 5 mm was developed [13]. Dynamic coherent illumination could be suited for this mesoscopic system. At depths shallower than 1mm, optical-resolution photoacoustic microscopic system have been developed [3]. Because of the optical focusing, the spatial extension of the source in the lateral dimension is constrained; however along the depth-of-focus computing the GMD could be useful to retrieve absorbers extended along the axial direction. Speckles grains are expected to be larger at imaging depth of microscopy. Additionally, it is required to minimize other sources of variability, such as detection noise, to measure a small variance in pixel values, e.g. by taking more measurements at the cost of acquisition time. Finally, for *in vivo* biomedical imaging, one may use the natural rapid decorrelation of perfused tissues, which is of the order of a few ms [14], to produce the temporally varying illumination patterns.

This work was funded by the European Research Council (grant number 278025), the Fondation Pierre-Gilles de Gennes pour la Recherche (grant number FPGG031) and the Plan Cancer 2009-2013 (project Gold Fever). O.K. acknowledges the support of the Marie Curie intra-European fellowship for career development (IEF).